\documentclass[a4paper,twoside]{article}
\newcommand{\affil}[1]{$^{\rm #1}$}
\usepackage[authoryear]{natbib}
\bibpunct{(}{)}{;}{a}{}{,}
\usepackage{graphicx}
\usepackage{amssymb}
\usepackage{amsmath} 
\date{} 
\newcommand{\mnras}{MNRAS}
\newcommand{\apj}{ApJ}
\newcommand{\prd}{PhRvD}
\newcommand{\apjs}{ApJS}
\newcommand{\aj}{AJ}
\title{\large\bf\flushleft The Influence of Evolving Dark Energy on Cosmology}
\author{\parbox{\textwidth}{\flushleft
\vspace{-0.5cm}
{\it Luke Barnes\affil{A,C}, Matthew J. Francis\affil{A}, 
Geraint F. Lewis\affil{A} and Eric V. Linder\affil{B}}\\
\vspace{0.4cm}
{\small \affil{A}\,School of Physics, University of Sydney, NSW 2006, Australia}\\ 
{\small \affil{B}\,Physics Division, Lawrence Berkeley National Laboratory, 
Berkeley, CA 94720, USA}\\ 
{\small \affil{C}\,Email: {\tt luke@physics.usyd.edu.au}}}}
\begin{document}
\twocolumn[
\begin{changemargin}{.8cm}{.5cm}
\begin{minipage}{.9\textwidth}
\vspace{-1cm}
\maketitle
%
%
\small{\bf Abstract: Observational evidence indicating that the expansion of the universe is accelerating has surprised cosmologists in recent years. Cosmological models have sought to explain this acceleration by incorporating `dark energy', of which the traditional cosmological constant is just one possible candidate. Several cosmological models involving an evolving equation of state of the dark energy have been proposed, as well as possible energy exchange to other components, such as dark matter. This paper summarises the forms of the most prominent models and discusses their implications for cosmology and astrophysics. Finally, this paper examines the current and future observational constraints on the nature of dark energy.}



\medskip
\medskip
\end{minipage}
\end{changemargin}
]
\small
\section{Introduction} \label{intro}
The observed acceleration of cosmic expansion is a landmark discovery in modern Cosmology. We don't know why the universe is accelerating, but it is reasonably clear that the explanation will require new fundamental physics---beyond our current understanding of particle physics, gravitation and the quantum vacuum. Cosmologists are now in the fortunate position of being able to ask the questions that will provide insight into ultimate physics beyond the reach of particle accelerators, as well as understanding the history and fate of the universe.

The present best fit cosmological model, known as the concordance model, combines data from many complementary sources including the WMAP and other observations of the CMB \citep{2003ApJS..148..175S}, large scale structure surveys such as the 2DFGRS \citep{2003MNRAS.346...78H} and the SDSS \citep{2003AJ....126.2081A} and supernovae data \citep{2003ApJ...598..102K,2004ApJ...607..665R}. The accelerated expansion of the universe is modelled via the `cosmological constant' ($\Lambda$), an entity that has been introduced, removed and re-introduced several times since Einstein's original introduction into his field equations. The cosmological constant has some special properties that make it a natural choice for inclusion into our models, however there are many plausible alternatives also permitted by current data. The purpose of this paper is to review cosmological model basics and then explore some general unheralded implications of the many cosmological models that utilise a generic `dark energy' rather than assuming \emph{a priori} a cosmological constant.

While there is good agreement between current observational data and cosmological models incorporating a non-zero cosmological constant, its introduction has not been without problems. The first is known as the coincidence problem. The energy density of matter decreases as the universe expands, proportional to the cube of the scale size of the universe. However, the energy density associated with the cosmological constant remains constant as the universe expands. Thus, that we should be observing at an epoch when the energy density of matter and the cosmological constant are of the same order of magnitude seems very unlikely.

The second problem is the infamous cosmological constant problem, discussed in detail in \citet{1989RvMP...61....1W} and \citet{2001LRR.....4....1C}. In short, the fact that the energy density of the cosmological constant is unchanged by the expansion of the universe suggests that it can be identified as a property of spacetime itself - a vacuum energy density. Such a zero-point energy can be calculated from quantum field theory. However the theory predicts that if the vacuum energy is not zero, then it should be a value that is 120 orders of magnitude greater than the energy density required for the cosmological constant.

\begin{figure}[htb]
\begin{center}
\includegraphics[scale=0.35, angle=-90]{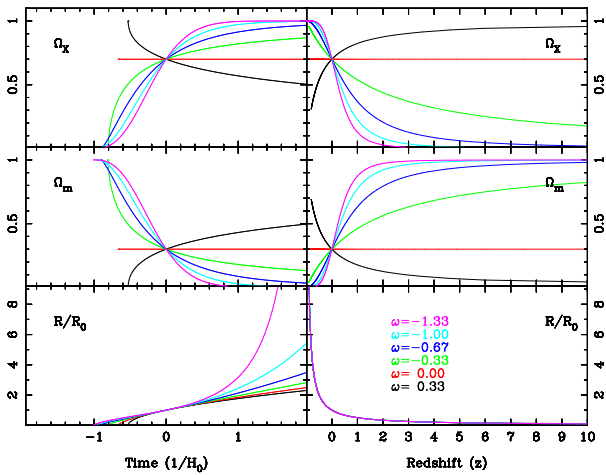}
\caption{Cosmological evolution for a constant equation of state: the top two panels show the energy parameter for the dark energy component $\Omega_{\Lambda}$, the middle two panels are for matter $\Omega_\textrm{m}$ and the bottom two plot the scale factor. The left panels show evolution with respect to time (in units of 1/$H_0$), while the right panels are with respect to redshift. The bottom right panel plots an identity (equation \eqref{eq:Rz}) and is shown for consistency.} \label{figure1}
\end{center}
\end{figure}

\begin{figure}[Htb]
\begin{center}
\includegraphics[scale=0.35, angle=-90]{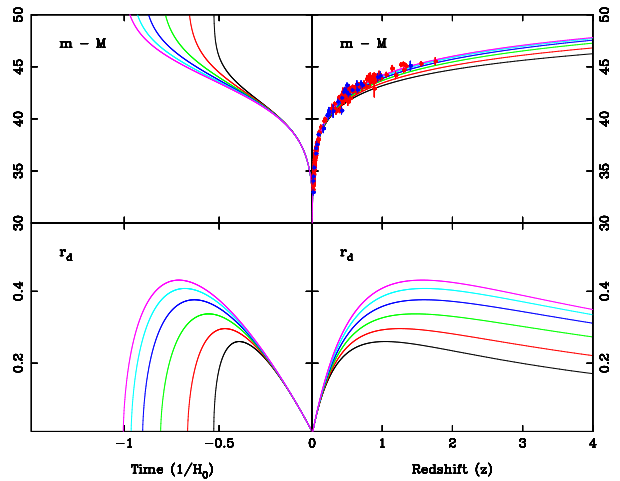}
\caption{Distance measurements for a constant equation of state: the top two panels show the effective magnitude as defined by equation \eqref{meff}, using the angular diameter distance in units of Mpc using an arbitrary value of h=0.72. 
The gold and silver data sets as in \cite{2004ApJ...607..665R} 
are shown in corresponding colours. The bottom two show the angular diameter distance in units of 1/$H_0$. The left side panels show evolution with respect to time (in units of 1/$H_0$), while the right panels are with respect to redshift.}\label{figure2}
\end{center}
\end{figure}

The problems with the cosmological constant, as well as the relative lack of observational constraints, have lead to a flurry of alternative explanations in recent years. A component that causes the expansion of the universe to accelerate is referred to as dark energy, with a cosmological constant being just one possibility. Another popular candidate is a primordial scalar field, or `quintessence'. This is the generic name for a time varying, spatially inhomogeneous component with negative pressure. The prime example is a scalar field $Q$ slowly rolling down a self-interaction potential $V(Q)$ \citep{1988PhRvD..37.3406R,1998ASPC..151...13S,2003RSPTA.361.2497S}. Unlike the cosmological constant, which is ascribed to vacuum energy, the quintessence field has no expected value and the 120 orders of magnitude problem can disappear. However, it is a much more ad hoc approach as there is no `natural' reason (besides possible connections with early universe inflation) to postulate the existence of a quintessence field, unlike the more comfortable vacuum energy interpretation of the cosmological constant. Many and varied quintessence models have been proposed \citep{2005astro.ph..2439B,2003PhRvD..67h3513C,2005astro.ph..4090B}. The evolution of the properties of the scalar field may solve the coincidence problem, as described in section \ref{coin}. There are also alternative models of gravity that seek to explain the observed data [for instance see \citep{2004PhRvD..70d3528C,2005PhRvD..71f3513C} and references therein] as well as explanations other than accelerated expansion. If distance supernovae were dimmed by some kind of `grey dust' then this would give the false impression cosmic acceleration, however see \citep{2004ApJ...607..665R} for a discussion of the problems with the grey dust model. More exotic dimming processes such as photon/axion mixing in magnetic fields \citep{2004astro.ph..9596C} have been proposed, though calcualtions show that this cannot alone account for the observed acceleration. Observational data are currently insufficient to cull the field of alternatives. The solution of this critical problem in modern cosmology will require a great deal of observational effort.

This paper examines the general form of several of the currently proposed models, including those involving energy transfer between dark energy components, and examines their influence on the past and future expansion of the Universe. Section \ref{back} gives a general background on cosmology, introducing the notation and important equations as well as outlining the current observational evidence. Section \ref{evol} introduces the equation of state and examines models where it is allowed to evolve with time. Section \ref{inter} introduces models with an interaction between dark energy density and matter density. Section \ref{cosm} discusses the implications of the models presented in sections \ref{evol} and \ref{inter}.

\section{Background} \label{back}
Modern theoretical cosmology is built on two pillars: the cosmological principle and general relativity. The cosmological principle states that we do not occupy a special position in the universe. This allows our observation that the universe is isotropic to be extrapolated to the global property of homogeneity. Homogeneity dramatically simplifies our cosmological theories, as it means that any parameter that describes the universe as a whole can only depend on time.

{}From the assumptions of homogeneity and isotropy, a metric can be derived that tells us how to measure distance and time in the universe. This metric is known as the Robertson-Walker metric (see \citet{Weinberg:1972} for more details), and has the line element:
\begin{figure}[Htb]
\begin{center}
\includegraphics[scale=0.35, angle=-90]{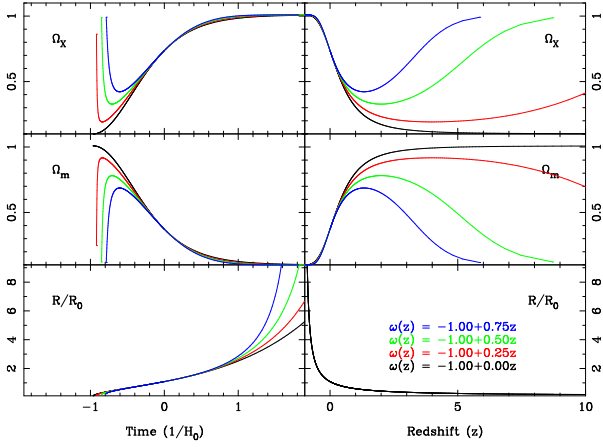}
\caption{Cosmological evolution for an equation of state linearly varying in redshift: the layout is the same as figure \ref{figure1}.  Note that this parameterisation breaks down for $z\gtrsim1$.} \label{figure3}
\end{center}
\end{figure}
\begin{figure}[Htb]
\begin{center}
\includegraphics[scale=0.35, angle=-90]{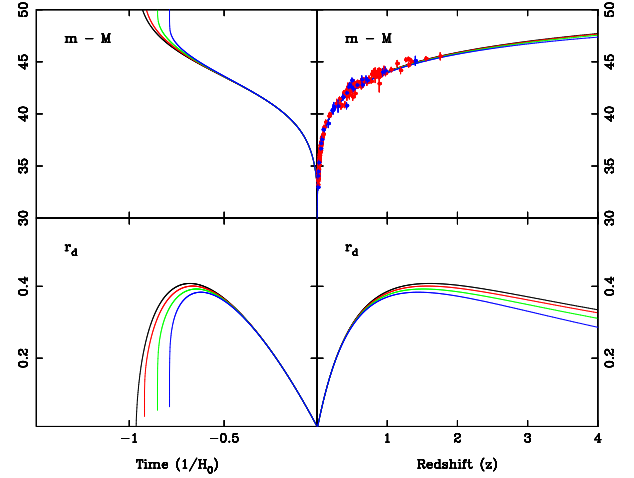}
\caption{Cosmological evolution for an equation of state linearly varying in redshift: the layout is the same as figure \ref{figure2}.  Note that this parameterisation breaks down for $z\gtrsim1$.} \label{figure4}
\end{center}
\end{figure}
\begin{multline} \label{eq:RW}
\textrm{d}s^2 = c^2 \textrm{d}t^2 - \\ R^2(t) \left(\textrm{d}\chi^2 +
S_k^2 \left(\textrm{d}\theta^2 + \sin^2\theta \textrm{d}\phi^2\right)\right)
\end{multline}
where $c$ is the speed of light (hereafter c = 1) and
\begin{gather}
S_k(\chi) = \begin{cases}
\sin \chi & \text{closed universe ($k$ = +1)}\\
\chi & \text{flat universe ($k$ = 0)}\\
\sinh \chi & \text{open universe ($k$ = -1)}\\
\end{cases}
\end{gather}
where $(\chi,\theta,\phi)$ are the spherical comoving coordinates, which for an object moving with the Hubble flow do not change as the universe expands. $R$ is the scale factor of the universe, which is the key prediction of any cosmological model. It contains information about evolution and fate of the universe, and is simply related to the observable redshift ($z$) by:
\begin{equation} \label{eq:Rz}
\frac{R(z)} {R_0} = \frac{1} {1+z}
\end{equation}
The subscript 0 refers to the present epoch. We also define the Hubble parameter $H$, which measures the rate of expansion of the universe:
\begin{equation} \label{eq:Hdef}
H \equiv \frac{\dot{R}} {R}
\end{equation}
Differentiating \eqref{eq:Rz} gives:
\begin{equation} \label{eq:dzdt}
H = - \frac{1} {1+z} \frac{\textrm{d}z} {\textrm{d}t}
\end{equation}

In the curved, expanding spacetime of the Robertson-Walker metric, not all methods of measuring distance are the same (see \citet{Linder:1997} for details and \citet{Hogg:1999ad} for a summary). The comoving radial distance ($r_\textrm{p}$) between the origin and $(\chi,\theta,\phi)$ is defined as a simultaneous ($\textrm{d}t = 0$) radial measurement ($\textrm{d}\theta = \textrm{d}\phi = 0$) at time $t_0$ and is given by:
\begin{equation} \label{eq:prop}
r_\textrm{p} = R_0 \chi
\end{equation}
Light travels along null geodesics, defined by $\textrm{d}s = 0$. Since we have placed ourselves at the centre of the coordinate system, light moves radially ($\textrm{d}\theta = \textrm{d}\phi = 0$). Noting that all the quantities are positive, the RW metric reduces to:
\begin{equation} \label{eq:chi1}
c\textrm{d}t = R(t) \textrm{d}\chi \qquad \Rightarrow \qquad \chi(t_{em}) = c\int_{t_{em}}^{t_0} \frac{\textrm{d}t'} {a(t')}
\end{equation}
Note that this is not simply multiplying the speed of light by the light travel time ($t_0-t_{em}$). To convert equation \eqref{eq:chi1} in terms of the observable redshift, we use \eqref{eq:dzdt}:
\begin{equation} \label{eq:chiz}
\chi(z) = c\int_0^z \frac{\textrm{d}z'} {H(z')}
\end{equation}
where we have used the fact that $t_{0}$ corresponds to $z=0$ and $t_{em}$ to $z$.

Proper transverse distance ($r_\textrm{m}$) and angular diameter distance ($r_\textrm{d}$) both relate the transverse proper size $l$ of an object at $\chi$ to the angular size ($\Delta\theta$). For $r_\textrm{m}$, all measurements are done simultaneously ($\textrm{d}t = 0$) at time $t_0$ giving:
\begin{equation} \label{eq:comt}
r_\textrm{m} = R_0 S_k(\chi)
\end{equation}
For $r_\textrm{d}$, measurements are done using photons which travel along geodesics ($\textrm{d}s = 0$). Thus, angular diameter distance to an object which emits light at time ($t_{em}$) as measured by an observer at $t_0$ is the proper transverse distance at $t_{em}$:
\begin{equation} \label{eq:ang}
r_\textrm{d} = R(t_{em}) S_k(\chi)
\end{equation}

Luminosity distance ($r_\textrm{l}$) to a source of intrinsic luminosity $L$ from an observer who measures its flux $S$ is defined by: $L = 4 \pi r_\textrm{l}^2 S$. The distances $r_\textrm{m}$, $r_\textrm{d}$ and $r_\textrm{l}$ are related by:
\begin{equation} \label{eq:dist}
r_\textrm{l} = (1+z) r_\textrm{m} = (1+z)^2 r_\textrm{d}
\end{equation}

The field equations of general relativity allow us to relate the metric of the universe \eqref{eq:RW} to its energy content. The result is the Friedmann equations:
\begin{subequations} \label{eq:frieds}
\begin{align}
H^2 &= \frac{8\pi G} {3} \rho - \frac{k} {R^2} \label{eq:fried} \\
\dot{\rho} &= -3H(\rho + p) \label{eq:cons}\\
\frac{\ddot{R}} {R} &= -\frac{4\pi G} {3} (\rho + 3p) \label{eq:accel}
\end{align}
\end{subequations}
where an overdot refers to differentiation with respect to time; $G$ is Newton's gravitational constant; $\rho$ is the total energy density; $p$ is the pressure; $H$ is the Hubble parameter; and $c = 1$ for convenience.

Equation \eqref{eq:fried} is known as the expansion equation, \eqref{eq:cons} is the adiabatic equation and equation \eqref{eq:accel} is the acceleration equation. Any of these equations can be derived using the other two, although all three equations are produced separately from the Einstein field equations of general relativity (and are related through the Bianchi identity). 

As is consistent with CMB data, including WMAP \citep{2003ApJS..148..175S}, we assume throughout that the universe is flat ($k=0$). We consider the universe to contain a number of components (labelled i), each with a corresponding pressure $p_\textrm{i}$ and density $\rho_\textrm{i}$ that contribute to the total $p = \sum_{\textrm{i}} p_\textrm{i}, \ \rho = \sum_{\textrm{i}} \rho_\textrm{i}$. Defining the critical density to be:
\begin{equation} \label{eq:crit}
\rho_\textrm{crit}(z) \equiv \frac{3H(z)^2}{8 \pi G}
\end{equation}
the energy components can then be described relative to $\rho_\textrm{crit}$:
\begin{equation} \label{eq:omegai}
\Omega_{\textrm{i}}(z) \equiv \frac{\rho_{\textrm{i}}(z)}{\rho_{\textrm{crit}}(z)}
\end{equation}
where $\Omega_{\textrm{i}}$ is the dimensionless density parameter for component i. Then, by putting $k=0$ into \eqref{eq:fried} the sum over all the density parameters is unity:
\begin{equation} \label{eq:nfried}
\Omega \equiv \sum_{\textrm{i}} \Omega_{\textrm{i}}(z) = 1
\end{equation}

Details of how the density parameters $\Omega_{\textrm{i}}(z)$ were calculated in this paper are contained in appendix \ref{numerics}.

\begin{figure}[Htb]
\begin{center}
\includegraphics[scale=0.35, angle=-90]{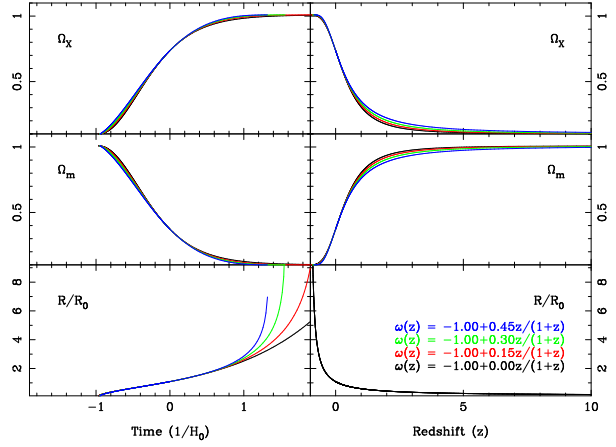}
\caption{Cosmological evolution for the linear parameterisation of the
equation of state detailed in section \ref{semlin}: the layout is the same as figure
\ref{figure1}.}\label{figure5}
\end{center}
\end{figure}

\begin{figure}[Htb]
\begin{center}
\includegraphics[scale=0.35, angle=-90]{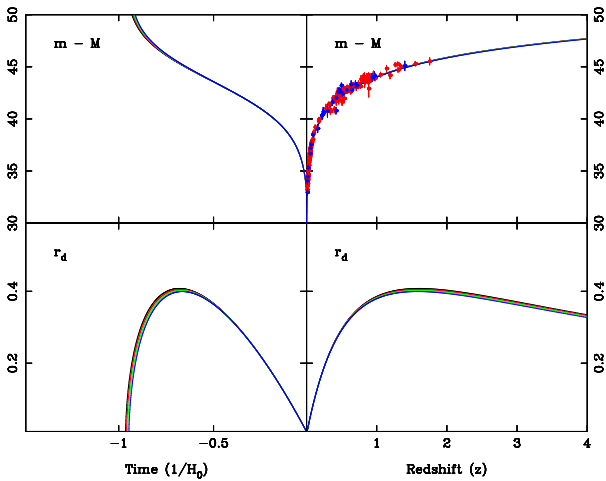}
\caption{Cosmological evolution for the linear parameterisation of the
equation of state of detailed in section \ref{semlin}: the layout is the same as figure
\ref{figure2}.} \label{figure6}
\end{center}
\end{figure}

As well as energy density, pressure also appears in the Friedmann equations. We commonly relate the pressure of a component to its energy density using an Equation of State (EOS):
\begin{equation} \label{eq:defw}
w_{\textrm{i}} \equiv \frac {p_{\textrm{i}}}{\rho_{\textrm{i}}}
\end{equation}
If the components of the universe are non-interacting (they do not exchange energy), then they will each satisfy their own adiabatic equation \eqref{eq:cons}:
\begin{equation} \label{eq:consw}
\dot{\rho_{\textrm{i}}}=-3H\rho_{\textrm{i}}(1+w_{\textrm{i}})
\end{equation}
For most of the familiar components of the universe, $w_\textrm{i}$ is constant: ordinary, non-relativistic matter is essentially pressureless and has $w_\textrm{m} = 0$, radiation has $w_\textrm{r} = 1/3$, a cosmological constant has $w_\Lambda = - 1$ \citep{1988A&A...206..175L}.

A cosmological model, capable of specifying $R(t)$ for all times, is described by specifying: the components of the universe, their equations of state $w_{\textrm{i}}$, their densities today $\Omega_{\textrm{i}}(0)$ and the present value of the Hubble parameter, $H_0$. Experimental uncertainty in $H_0$ is often 
written in terms of the dimensionless $h$, defined by $H_0 = 100h \mathrm{Mpc^{-1} km\, s^{-1}}$. The dependence of quantities on $H_0$ can then be made explicit, e.g. $\Omega_{\textrm{i}}h^{2}$. The value of h was measured by the Hubble Key Project team to be $h=0.72\pm8$ \citep{2001ApJ...553...47F}.

\subsection{Observational Evidence}
Several independent lines of evidence lead to the conclusion that the expansion of the universe is accelerating. Recent observations of the CMB from the WMAP satellite have confirmed that the universe appears to be flat, with $\Omega = 1.02 \pm 0.02$ \citep{2003ApJS..148..175S}. The power spectrum of galaxy distributions on the other hand points to $\Omega_{\textrm{m}} \simeq 0.3$ \citep{2002MNRAS.337.1068P}. A key third measurement is the magnitude-redshift relation of supernovae type 1a (hereafter SN1a), see \citet{2004ApJ...607..665R} for a compilation of recent data. These measurements give strong evidence that the expansion of the universe is accelerating, and has been accelerating since $z \sim 1$. The concordance model accounts for the missing energy density needed to make the universe flat via a cosmological constant ($w = -1$), however this is by no means the only permitted form of dark energy. A new generation of cosmological experiments, including the proposed SuperNova Acelleration Probe (SNAP) \citep{Aldering:2004} are expected to measure more supernovae at higher redshifts. This new data will be crucial in discriminating between cosmological models.

SN1a are very bright standard candles, making them excellent cosmological probes. The distance modulus, once the appropriate K-correction has been applied is given by:
\begin{equation} \label{meff}
m-M = 25 + 5\log(H_0 r_{\textrm{l}})
\end{equation}
where $r_{\textrm{l}}$ is defined in section \ref{back} and is presented here in units of Mpc. All three distances previously defined are model dependent and hence the measurement of the magnitude and redshift of SN1a can be used to discriminate between models. The methods used to compute distances are described in appendix \ref{numerics}. The results are presented in the even numbered figures in this paper. In these figures SN1a data compiled in \cite{2004ApJ...607..665R} have been added in order to illustrate how similar all the models are in the region $z<0.5$ where the data sit. The SNAP probe \citep{Aldering:2004} is anticipated to reach $z \simeq 1.5 $, where the models can be seen to diverge much more than for the current data set (see even numbered figures), illustrating the much greater constraints on the models that will be possible.
\begin{figure}[Htb]
\begin{center}
\includegraphics[scale=0.35, angle=-90]{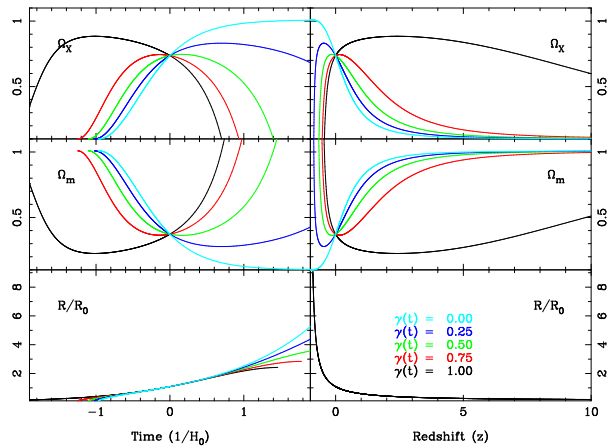}
\caption{Cosmological evolution for a constant interaction between matter and dark energy: the layout is the same as figure \ref{figure1}.}\label{figure7}
\end{center}
\end{figure}

\begin{figure}[Htb]
\begin{center}
\includegraphics[scale=0.35, angle=-90]{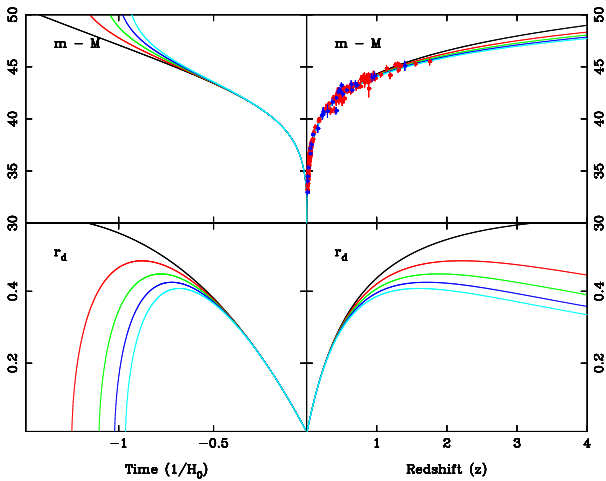}
\caption{Cosmological evolution for a constant interaction between matter and dark energy: the layout is the same as figure \ref{figure2}}\label{figure8}
\end{center}
\end{figure}

The data described above lead to a cosmological model where
\begin{equation*}
k = 0 \quad (\textrm{The universe is flat})
\end{equation*}
\begin{equation*}
h \simeq 0.7 \quad
\Omega_\textrm{m0} \simeq 0.3 \quad
\Omega_{\Lambda0} \simeq 0.7 
\end{equation*}
Note however that this model makes the assumption that the dark energy is a true cosmological constant. This assumption is permitted by existing data, but not strongly favored over other possibilities. Despite this untested assumption this model is used widely. In this paper the universe is assumed to be flat with $\Omega_{\textrm{m0}}=0.3$ and $\Omega_{\textrm{X0}}=0.7$ as in the concordance model. Note that the subscript X is used to denote dark energy, leaving $\Lambda$ to denote the specific case of a cosmological constant. $H_{0}$ is set to unity, which is equivalent to measuring time in units of $H_{0}^{-1}$.

\section{Equation of State of Dark Energy} \label{evol}
In this section the effect of different forms of the dark energy EOS (as defined by \eqref{eq:defw}) on the expansion history and distance measurements are examined. We take a phenomenological view of the EOS, rather than speculating on the physical processes behind any particular form. This is a common approach given the mysterious nature of dark energy and current data being unable to constrain complex models. It was shown in \citet{2003MNRAS.346..573L} that any term other than matter in \eqref{eq:fried} can be modelled by an effective equation of state of dark energy, regardless of the physical origin of the term. If we write \eqref{eq:fried} as
\begin{equation}
H^2 = \Omega_{\textrm{m}}R^{-3} + \delta H^2
\end{equation}
with $H_0$ set to unity, then by comparison to the case of dark energy with some equation of state $w$
\begin{equation}
H^2 = \Omega_{\textrm{m}}R^{-3} + (1- \Omega_{\textrm{m}})R^{-3(1+ w)}
\label{eq:h2w}
\end{equation}
and
\begin{equation}
w = -1 -\frac{1}{3}\frac{d(\ln \delta H^2)}{d(\ln R)}
\end{equation}
For any arbitrary mechanism that causes $\delta H^2$, there is a corresponding function $w$ (not necessarily constant, as in \eqref{eq:h2w}) that can be calculated. This allows a wide variety of different models and mechanisms to be compared within a single parameter space (see \citet{2004PhRvD..70f1302L}).

\begin{figure}[Htb]
\begin{center}
\includegraphics[scale=0.35, angle=-90]{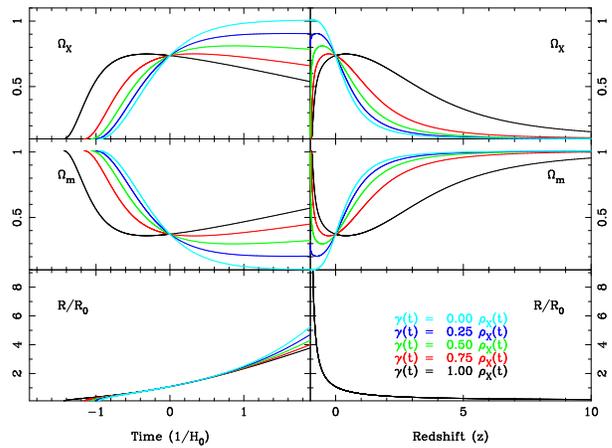}
\caption{Cosmological evolution for a decaying dark energy model: the layout is the same as figure \ref{figure1}.}\label{figure9}
\end{center}
\end{figure}

\begin{figure}[Htb]
\begin{center}
\includegraphics[scale=0.35, angle=-90]{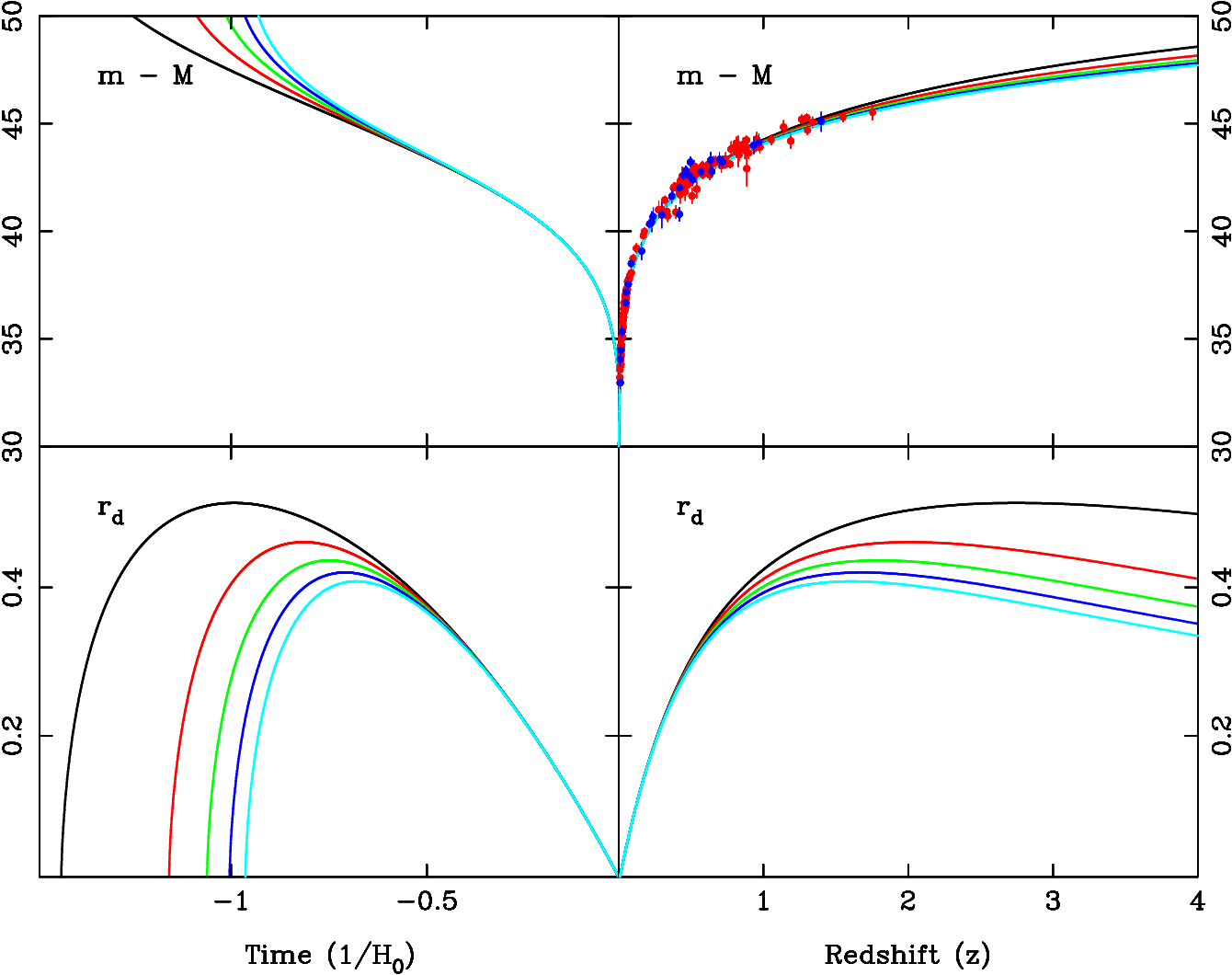}
\caption{Cosmological evolution for a decaying dark energy model: the layout is the same as figure \ref{figure2}.}\label{figure10}
\end{center}
\end{figure}

\subsection{Constant Equation of State}
Current data are insufficient to constrain $w_{\textrm{X}}$ to more than one parameter \citep{2004PhRvD..70l3516L}; in other words, a constant $w_{\textrm{X}}$. Using CMB, LSS and SN1a data, \citet{2004ApJ...606..654W} found the constraint on a constant dark energy EOS to be $-1.24 < w_{\textrm{X}} < -0.74$. We present a wider range of possibilities than this, in order to exaggerate the effects for clarity. When $w_\textrm{i}$ is constant, equation \eqref{eq:consw} can be easily integrated to give $\rho_\textrm{i} = \rho_\textrm{i} (z)$ using equation \eqref{eq:Rz}:
\begin{equation} \label{eq:rhoz1}
\rho_\textrm{i} (z) = \rho_{\textrm{i}}(0) (1+z)^{3(1+ w_\textrm{i})}
\end{equation}
Equation \eqref{eq:rhoz1} gives the familiar results that, as we look back into the universe, $\rho_\textrm{m} \propto(1+z)^3$, $\rho_\textrm{r} \propto (1+z)^4$, whilst $\rho_\Lambda$ remains constant as the universe expands. From \eqref{eq:rhoz1} we can see that as the universe ages, the component with the most negative $w$ will come to dominate the total energy of the universe. Consider equation \eqref{eq:accel} in the case where the component with the most negative $w$ has come to dominate the total energy of the universe. If $w < -\frac{1}{3}$, then $\ddot{R}$ will be positive and the universe will accelerate. We can therefore divide components into accelerating ($w < -\frac{1}{3}$) and decelerating ($w >\frac{1}{3}$). If $w = -\frac{1}{3}$ then the expansion coasts: $\dot{a} = $constant.

Thus, if the universe has any accelerating components, the universe will eventually begin to accelerate, independent of its geometry \footnote{The exception is where the energy density in the decelerating components is high enough that the universe begins to contract before the accelerating components come to dominate the total energy density.}. Only if all the components are decelerating will the universe decelerate, and its fate be decided by its geometry.

Things get interesting when $w < -1$. If we consider an equivalent form of equation \eqref{eq:rhoz1}:
\begin{equation*}
\rho_\textrm{i} (R) = \rho_{\textrm{i0}} \left( \frac{R}{R_0} \right)^{-3(1+ w_\textrm{i})}
\end{equation*}
we see that $w < -1$ implies that the energy density of the component will increase as the universe expands. Such a component has been dubbed ``phantom energy" and its consequences have been studied in \citet{2003PhRvL..91g1301C}. In short, phantom energy causes the scale size of the universe to approach infinity in finite time. The resulting ``$R = \infty$'' singularity has been dubbed ``the big rip''. Before the end, the expansion of the universe will overpower all other forces - first gravity, as the Milky Way, the Solar System and finally the earth are pulled apart, followed by weak, EM and strong nuclear forces as matter is torn into its components.

In figures \ref{figure1} and \ref{figure2} we consider models identical to the concordance model (matter + dark energy) except that we allow the equation of state for dark energy to vary. Figure \ref{figure1} shows the expansion history and density parameters while \ref{figure2} shows the measurable distance properties with the current SN1a data overlaid.

The pink model ($w = -4/3$) is an example of a phantom energy model. The ``big rip'' behaviour as time increases can be seen in the bottom left panel of figure \ref{figure1}. The light blue model ($w = -1$) is the concordance model. For the light green model ($w = -0.33$), we see from equation \eqref{eq:accel} that as dark energy begins to dominate, the acceleration goes to zero. Thus, $R(t)$ approaches a linear function as time increases. The red model ($w = 0$) is the Einstein-de-Sitter model (flat, matter only), which decelerates as time increases but never turns around and begins to contract. The black model ($w = 0.33$) contains a radiation component whose energy density falls quickly as the universe expands.

\subsection{Evolving Equation of State}
While current data are insufficient to constrain $w_{\textrm{X}}(z)$ to anymore than one parameter \citep{2004PhRvD..70l3516L}, next generation data, such as the Planck Surveyor \citep{2004AdSpR..34..491T} and the SNAP satellite \citep{Aldering:2004} will be able to constrain possible evolution of $w_{\textrm{X}}(z)$. This section presents two different first order parameterisations of $w_{\textrm{X}}(z)$ and demonstrates the effects of several values of the parameters. Since an evolving EOS will only be considered for the dark energy component ($\textrm{i} = \textrm{X}$), we omit the subscript on $w$ in this section. Again using \eqref{eq:dzdt} and \eqref{eq:cons}  
\begin{equation} \label{eq:rhoz2}
\rho_\textrm{X} (z) = \rho_{\textrm{X}}(0) \ \textrm{e}^{3 \int_0^z \frac{1+\omega (z')} {1+z'} dz'}
\end{equation}

While a two parameter evolving equation of state $w (z)$ could take many forms, two are most common in the literature.

\subsubsection{Old Linear Parameterisation}
The simplest first order expansion of the equation of state gives $w (z) = w_0 + w_1 z$. While obsolete (see \ref{semlin}), this form has been used widely in the past. In this case, equation \eqref{eq:rhoz2} becomes:
\begin{equation} \label{eq:rhozlin}
\rho_\textrm{X} (z) = \rho_{\textrm{X}} (0) \ (1+z)^{3(1+\omega_0 - \omega_1)} \textrm{e}^{3\omega_1 z}
\end{equation}

In figures \ref{figure3} and \ref{figure4} we show the effects of a modified concordance model by using a linear parameterisation of the dark energy equation of state.

 The first thing to note in Figure \ref{figure3} is the bizarre behaviour of the energy densities at early time (i.e. large redshift). This shows the inadequacy of this simple parameterisation: it is inappropriate for $z \gtrsim 1$ as $w(z)$ is unbounded as $z \to \infty$ (as is $\rho_X(z)$ for $w_1>0$). 

As we model into the future, further problems arise. For a universe that is always expanding, forward modelling is done by allowing $z \to -1$. In this case, $w (z) \to w_0 - w_1$. If $w_0 - w_1 < -1$, then eventually $w < -1$ and the universe will end in a ``big rip''.

\subsubsection{Linear Parameterisation} \label{semlin}
In \citet{2003PhRvL..90i1301L}, a parameterisation is used of the form: $w (z) = w_0+w_1(1-R/R_0)=w_0 + w_1 z / (1 + z)$. Then the equation for the density becomes:
\begin{equation} \label{eq:rhozeric}
\rho_\textrm{X} (z) = \rho_{\textrm{X}}(0) \ (1+z)^{3(1+ w_0 + w_1)} \textrm{e}^{-3 w_1\frac{z}{1+z}}
\end{equation}

The advantages of this parameterisation are given in \citet{2003PhRvL..90i1301L}, in short the old parameterisation clearly becomes problematic at high $z$, while this parameterisation has a bounded behaviour as $z \to \infty$. The results of several values of the parameters in this model are shown in figures \ref{figure5} and \ref{figure6}.

The energy densities for the parameters shown in figure \ref{figure5} are remarkably similar throughout the lifetime of the universe. The past history of $R(t)$, too, is very similar. However, the parameterisation is unsuitable for modeling the future, 
$z\to-1$, since $w(z)$ becomes unbounded.  Note by contrast that a 
parameterisation linear in redshift is unbounded in the past -- where all 
the data is!  Models such as the ``e-fold'' case of \citet{2005astro.ph..5330L} and 
the ``kink'' case of \citet{2003PhRvL..90i1303C} can smoothly handle both the past 
and future.

\section{Interacting Components} \label{inter}

In the previous section we assumed that dark energy does not interact with matter or radiation. However, as we have no knowledge of the microphysics of dark energy, it is worth considering what the effects of interacting dark energy might be. Studies of such models are ongoing, see \citep{2005gr.qc.....5056Z} for a recent example. Much of this work has been presented in Physics rather than astronomical journals, as the interest centre of the nature of the fields involved in the interaction. Here we focus the general form of such models and highlight the consuqunces that may be of interest to astronomers and cosmologists.

The Friedmann equations do not restrict energy components to be non-interacting and can be easily modified to model an energy exchange. Equation \eqref{eq:consw} is true in the case where the component i is not exchanging energy with any other components. We now return to equation \eqref{eq:cons} and relax this assumption. Consider the case where there are only two components in the universe: pressureless matter and dark energy. Then \eqref{eq:cons} can be split into its components \citep{2005astro.ph..2034S}:

\begin{equation} \label{eq:cons2}
(\dot{\rho}_{\textrm{m}} + 3H\rho_{\textrm{m}}) + ( \dot{\rho}_{\textrm{X}} + 3H\rho_{\textrm{X}} (1+ w_{\textrm{X}})) = 0
\end{equation}

At this point, a phenomenological parameter $\gamma$ can be introduced to represent the interaction between dark energy and matter. Then:
\begin{align}
\dot{\rho}_{\textrm{m}} + 3H\rho_{\textrm{m}} &= \gamma \label{eq:consm} \\
\dot{\rho}_{\textrm{X}} + 3H\rho_{\textrm{X}} (1+ w_{\textrm{X}}) &= -\gamma \label{eq:consX} \end{align}

$\gamma$ has units of energy per unit time per unit volume, and represents the rate at which energy is transferred from the dark energy component to matter ($\gamma > 0$) or vice versa ($\gamma < 0$).

We cannot simply integrate \eqref{eq:consm} and \eqref{eq:consX} in the same way we did to obtain equation \eqref{eq:rhoz2}. To find $\rho_{\textrm{m}}$, $\rho_{\textrm{X}}$, and $R(t)$ we need to solve \eqref{eq:consm} and \eqref{eq:consX} simultaneously with the expansion equation \eqref{eq:fried}. This represents a coupled set of (potentially) non-linear ODE's which are solved numerically (see appendix \ref{numerics}).

Note that in this section we take $w_{\textrm{X}}=-1$ for all the models, 
though this is not a cosmological constant.  We do this to isolate the 
effects of the interaction on the evolution.  Energy transfer between matter and other forms of dark energy could also be computed, however we are interested in examining the broad consequences of energy transfer, which will be qualitatively similar for other forms of dark energy.  Future work will generalize this; see also 
\cite{2005astro.ph..7263L}.

\subsection{Constant Interaction}
The first form of interaction considered was a constant transfer of energy, $\gamma= \textrm{constant}$. We considered only the case where $\gamma$ is non-negative, where energy is being transferred from dark energy into matter. Concordance models with this modification are shown in Figures \ref{figure7} and \ref{figure8}.

In these models, the energy density in dark matter peaks before decreasing and eventually becomes negative, at which point the models were stopped. The density of a component crossing through zero violates either the Big Bang condition or the continuity equation \citep{2004PhRvD..70f1302L} and therefore a simple constant $\gamma$, while it may give insight into the qualitative effects of energy transfer, cannot model an interaction suitably over the full range of cosmic time.

\subsection{Dark Energy Decay}
To avoid the problem of negative dark energy density, we considered a decaying dark energy model with a similiar form as in \citet{1985PhRvD..31.1212T}. Here $\gamma$ has the form:
\begin{equation} \label{eq:decay}
\gamma = \lambda \rho_X
\end{equation}

where $\lambda$ is a decay constant. This situation is analogous to radioactive decay, with dark energy decaying into matter. This model is shown in figures \ref{figure9} and \ref{figure10}

In these models, the dark energy density peaks before decreasing asymptotically to zero. Thus, dark energy only dominates the energy density of the universe for a finite period of time, and in the future will be negligible, just as it was in the past. To illustrate this point, we can calculate the deceleration parameter $q(t)$ for these models. The deceleration parameter is given by:
\begin{align} \label{eq:qt}
q(t) &\equiv -\left( \frac{\ddot{R}} {R} \right) \left( \frac{R} {\dot{R}} \right)^2 \\
& = \left( \frac{1} {2} \right) \left( \frac{\rho+3p} {\rho} \right)
\end{align}
by equations \eqref{eq:fried} and \eqref{eq:accel} with $k = 0$. $q(t)$ is defined so that a decelerating universe has $q>0$. An Einstein-de-Sitter universe has $q = 1/2$, a cosmological constant dominated universe has $q = -1$, and, in general, a universe dominated by an energy component with equation of state $w$ has $q = (1+3w)/2$.

\begin{figure}[Htb]
\begin{center}
\includegraphics[scale=0.25]{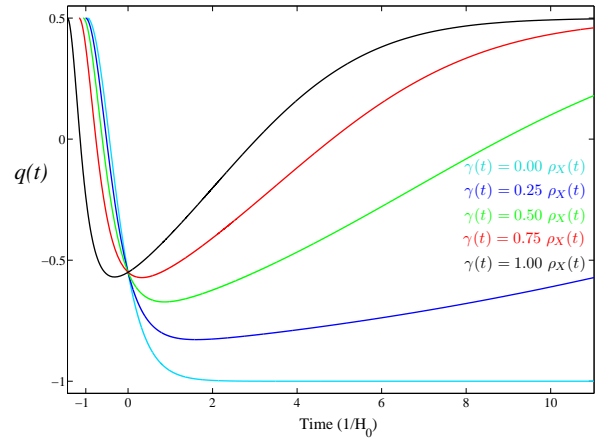}
\caption{The deceleration parameter for a decaying dark energy model: The light blue model is the concordance model, where the universe has started accelerating in the recent past and will always accelerate in the future. By contrast, the models with decaying dark energy only experience a finite period of accelerated expansion, beginning and ending like an Einstein-de-Sitter universe.} \label{qfig}
\end{center}
\end{figure}

A plot of $q(t)$ is shown in figure \ref{qfig}. In the early universe, matter dominates and the universe is decelerating. The energy density of matter quickly drops as the universe expands, and the dominance of dark energy causes the universe to accelerate. However, the decay of dark energy into matter eventually sees matter dominate and the universe decelerates again. Such a model would begin and end like an Einstein-de-Sitter universe, but would experience an epoch of accelerated expansion. 
\subsection{The Thermodynamics of Dark Energy}
This paper has approached dark energy phenomonelogically, parametrising its effects on the expansion history of the universe without worrying about their physical interpretation. However, the ultimate goal of the study of dark energy is physical understanding, where dark energy models are motivated by a deeper understanding of particle physics. Of particular interest is the physical meaning of the interaction parameter $\gamma$.

To this end, consider the adiabatic equation \eqref{eq:cons}. It can can be derived from the first law of thermodynamics for a closed, constant entropy system:
\begin{equation} \label{eq:thermo}
\textrm{d}E = \textrm{d}(\rho V)=-p\textrm{d}V
\end{equation}

In our interacting models, we assumed that energy was being exchanged between components. Thus we consider the  first law of thermodynamics as it applies to two open systems of $N_i$ particles (or quanta), which are exchanging particles. We introduce the chemical potential $\mu$:
\begin{equation} \label{eq:chempot}
\textrm{d}E_i=-p_i\textrm{d}V+\mu_i \textrm{d}N_i
\end{equation}

Comparing this equation to equations \eqref{eq:consm} and \eqref{eq:consX}, we find that our phenomenological parameter $\gamma$ can be written as:
\begin{equation} \label{eq:gamma}
\gamma = \frac{\mu_m} {R^3} \frac{\textrm{d}N_m} {\textrm{d}t} = -\frac{\mu_X} {R^3} \frac{\textrm{d}N_X} {\textrm{d}t}
\end{equation}
As would be expected, $\gamma$ is proportional to the rate of exchange of particles per unit volume, and can be considered the rate of exchange of energy density.
 
We now consider the relationship between $\gamma$ and the dark energy equation of state. By taking gamma over to the left hand side, equations \eqref{eq:consm} and \eqref{eq:consX} become:
\begin{align}
\dot{\rho}_{\textrm{m}} + 3H\rho_{\textrm{m}} (1 - \frac{\gamma} {3H\rho_{\textrm{m}}}) &= 0 \label{eq:wm} \\
\dot{\rho}_{\textrm{X}} + 3H\rho_{\textrm{X}} (1+ w_{\textrm{X}} + \frac{\gamma} {3H\rho_{\textrm{X}}}) &= 0 \label{eq:wX}
\end{align}

This suggests that we can simply consider interaction as a particular example of evolving equations of state for dark energy and matter. That is, we can define the effective dark energy equation of state to be: $w_{\textrm{X,eff}} = w_{\textrm{X}}+ \gamma / 3H\rho_{\textrm{X}}$ and the effective equation of state of matter to be: $w_{\textrm{m,eff}} = - \gamma / 3H\rho_{\textrm{m}}$.

But is it a good idea to consider interacting component models to be a special case of evolving equations of state? Phenomenologically, the answer is yes. In studying the expansion history of the universe, this strategy will allow us to examine both kinds of models in a common parameter space, knowing that it has no effect on the solution of the Friedmann equations for $a(t)$.

However, from the standpoint of physical understanding the answer is no. There is a marked difference between physical processes that create matter and those that change its equation of state. Matter with a non-zero equation of state has pressure comparable to its energy density. Pressure is related to momentum, so matter with a non-zero equation of state is matter moving at relativistic speeds\footnote{Recall $E^2 = (pc)^2 + (mc^2)^2$. The $(mc)^2$ term is so large for matter that only matter with $v \to c$ has non-negligible momentum.}. Thus reducing interacting components to a special case of evolving equations of state blurs the distinction between creating pressureless matter and accelerating matter to relativistic speeds. While this has no effect on expansion history ($a(t)$), it will undoubtedly affect structure history, i.e. the growth of inhomogeneities.

One final point can be made from thermodynamic considerations of dark energy. Conservation of energy in General Relativity is not well understood. Some are happy to say that energy is simply not conserved \citep{1995ApJ...446...63H}; Einstein introduced the gravitational stress energy tensor ($t^{\mu \nu}$) to represent the energy of the gravitational field to restore energy conservation \citep{einstein}, but this leads to problems because it is a pseudo-tensor: non-localizable, non-unique and non-covariant \citep{1973grav.book.....M}. Other approaches include altering Einstein's equations to include $t^{\mu \nu}$, \citep{yilmaz}, sparking refutation and counter-refutation \citep{Misner:1995na,Alley:1995sw}, and using path groups to formulate Gaussian flux integrals in curved spacetime \citep{2004PhLA..328..261M}.

In light of this we must be careful in applying the first law of thermodynamics in a cosmological context. We previously assumed that energy leaving the dark energy component will appear in the matter component. This is equivalent to saying that the universe as a whole is a closed system:
\begin{equation} \label{eq:consN}
\mu \textrm{d}N =  \mu_m \textrm{d}N_m + \mu_X \textrm{d}N_X = 0
\end{equation}
This seems very reasonable, but the same does not apply to the work term ($p\textrm{d}V \neq 0$). The work done by pressure in an expanding universe does not reappear in any other form, which is generally interpreted as non-conservation of energy in an expanding universe \citep{1995ApJ...446...63H}. Thus, we should keep in mind that it may be possible to construct plausible dark energy models where \eqref{eq:consN} does not hold.

\section{Cosmological Implications} \label{cosm}
We have discussed cosmological models that explore alternatives to a cosmological constant in explaining the accelerated expansion of the universe. The testing of dark energy models by observational data is a crucial challenge for cosmology in the coming years. We now outline some consequences of the range of presently allowed models.

\subsection{Age of the Universe} Dynamical dark energy models will give a different expansion history $R(t)$ than the concordance model for the same present day values of the density parameters. While constraints from analysis of the CMB and other measures, such as galaxy distributions, need to be carefully considered \citep[see][]{2005PhRvD..71f3523O}, dynamical dark energy may indicate that the age of the universe is different than expected from the concordance model. This highlights what should be obvious: the age of the universe is model dependent. Care must be taken in this area, for instance in \citet{2005astro.ph..2439B}, the parameters of a dark energy decay model are constrained by assuming the age of the universe which is found using the concordance model, a different model from that being constrained. Self-consistency in treatment of dark energy cosmologies is essential. This demonstrates the way in which the figure of $13.7 \pm 0.2$ Gyrs has become broadly applied: this figure is often quoted without acknowledging the assumption of a cosmological constant. To illustrate the increased uncertainty in the age of the universe when this assumption is relaxed, we find, using the zeroth order constraint on the dark energy EOS, $w_{\textrm{X}}$ of $-1.24 < w_{\textrm{X}} < -0.74$ as found in \citet{2004ApJ...606..654W}, that the derived age of the universe lies between $12.8$ and $13.9$ Gyrs, using $h=0.7, w_{\textrm{m}}=0.3, w_{\textrm{X}}=0.7$ as throughout this paper. This is not a rigorous analysis, as degeneracies between $w_{\textrm{X}}$ and $\Omega_{m}$ are likely to narrow this constraint, but does show the order of the increased uncertainty. Future work will determine a more precise value. 

\subsection{Coincidence Problem} \label{coin} It has been proposed that allowing a transfer of energy from dark energy to dark matter may help resolve the coincidence problem \citep[see][]{2005JCAP...03..002C, 2000PhRvD..62f3508C, 2001PhRvD..63j3508C, 2005astro.ph..3075Z, 2001PhLB..521..133Z}. In figures \ref{figure7} and \ref{figure9} the density parameters have a similar value for a far greater range of cosmic time than in the non-interacting case. With the coincidence problem in mind, the form of the interaction can be made to keep closer values of the parameters over a greater range of cosmic time \citep[e.g.][]{2003PhRvD..67h3513C}. Such a `solution' to the coincidence problem allows the fine-tuning of model parameters to be relaxed far more than in the concordance model, however it requires that the interaction term in the model itself be tuned. \citet{2004astro.ph..8495W} have shown that a cosmological constant decaying into matter is not likely to be permitted by even the current data set for any form of the interaction, though other forms of decaying dark energy (i.e. with $w_{\textrm{X}} \neq -1$) are not so constrained by their arguments. In any case further knowledge of dark energy dynamics will be a key to furthering our understanding of the apparent coincidence observed.

\subsection{Astrophysical Consequences}
Many astrophysical models, such as CDM models of galaxy halos, require knowledge of the cosmological parameters, $ \Omega_{\textrm{m}}, \ \Omega_{\textrm{X}}$ and $H_0$, usually assumed at present to be those of the concordance model. However, a universe with dynamic dark energy will not be satisfactorily described at all epochs by any values of these parameters alone. The effects of dark energy/dark matter interactions on CDM galaxy halo models is discussed in \citet{2004A&A...415..813G} and references therein; \citet{amendola} in particular has investigated several aspects of interaction. Future observations on cosmological scales that could constrain the cosmological model may greatly improve the understanding of galaxy formation and dynamics and vice versa. Note however that the models presented in this paper, as in most cosmological models, assume homogeneity. This assumption is supported on very large scales by galaxy surveys. However, while dark energy must be smooth over large scales [or else it would show up as an additional $\Omega_m$ in galaxy clustering surveys \citep[see][]{2001LRR.....4....1C}], dark matter is clumpy. Our phenomenological interaction between the dark sectors gives no clue as to how the smoothness of dark energy evolves to clumpy dark matter as this interaction proceeds.  This is an important outstanding issue. 

\section*{Acknowledgments}

LB is supported by University of Sydney School of Physics and Faculty of science scholarships. MF is supported by University of Sydney Faculty of Science UPA scholarship.

\appendix
\section{Numerical Model} \label{numerics}
The numerical model used to generate the results of this paper consisted of a system of first order, coupled ODE's which were then integrated in a Runge-Kutta scheme. The solution of the Friedmann equations can be tackled in two ways:

We can solve for $R$ in terms of $t$ by solving the equations:
\begin{gather}
H^2 = \frac{8\pi G} {3} (\rho_m + \rho_X) - \frac{k} {R^2}\\
\dot{\rho}_{\textrm{m}} = -3H\rho_{\textrm{m}} + \gamma \\
\dot{\rho}_{\textrm{X}} = -3H\rho_{\textrm{X}} (1+ w_{\textrm{X}}(z)) - \gamma
\end{gather}

The alternative is to solve for the energy density of the components in terms of $z$, using the equations:
\begin{gather}
\frac{\textrm{d}\rho_m} {\textrm{d}z} = \frac{3\rho_m} {1+z} - \frac{\gamma} {(1+z) H(z)} \\
\frac{\textrm{d}\rho_X} {\textrm{d}z} = \frac{3\rho_X(1+w_X(z))} {1+z} + \frac{\gamma} {(1+z) H(z)} \\
H(z)^2 = \frac{8\pi G} {3} (\rho_m + \rho_X) - \frac{k (1+z)^2} {R_0^2}
\end{gather}
Then, $t$ as function of $z$ could be obtained by integrating:
\begin{equation}
\textrm{d}t = - \frac{\textrm{d}z} {(1+z) H(z)} \\
\end{equation}

The present is $z=0$, and modelling into the past is done by allowing $z \to \infty$, and modelling into the future is done by allowing $z \to -1$. $R(t)$ can then be recovered from $z(t)$.

This second approach makes it easier to calculate the angular diameter and related distances. This could be done using the equations in section \ref{back}. However, the equations for angular diameter distance have been developed in the case where the universe is ``clumpy'' i.e. where it is only homogeneous and isotropic on average. The resultant equation is the Dyer-Roeder equation, as generalised in \citet{1988A&A...206..190L}:
\begin{multline}
\ddot{r}_\textrm{d} + \left( \frac{3+q(z)} {1+z} \right) \dot{r}_\textrm{d} + \\ \left( \frac{3} {2(1+z)} \sum_i (1+3w_i(z)) \alpha_i(z) \Omega_i(z) \right) r_\textrm{d} = 0
\end{multline}
where $q(z)$ is the deceleration parameter, and the initial conditions are:
\begin{gather}
r_\textrm{d}(z_0,z_0) = 0 \\
\frac{\textrm{d}r(z_0,z)} {\textrm{d}z} \Bigr\rvert_{z=z_0} = \frac{H_0}{H(z_0) (1+z_0)}
\end{gather}
where we have deliberately not set $z_0=0$ so that angular diameter distance can be calculated for observers at any redshift, as is common in situations involving gravitational lensing. The clumpiness parameter ($\alpha_i$), defined to be the ratio of the amount of component $i$ smoothly distributed to the total. It is set to 1 in this paper, which gives the same solutions for angular diameter distance as section \ref{back}. Our codes can relax this condition, and future papers may explore the effect of inhomogeneities in dark energy models.
\end{document}